\documentclass[preprint,12pt]{elsarticle}
\usepackage{multirow} 
\usepackage{stfloats} 
\usepackage{graphicx} 
\usepackage{bbding}
\usepackage{hyperref}
\usepackage{algorithm}
\usepackage{algpseudocode}
\usepackage{amsmath}
\usepackage{amssymb}
\usepackage{booktabs}
\usepackage{graphicx} 		
\usepackage{subfigure}		
\usepackage{xcolor}
\definecolor{mycolor}{RGB}{239, 158, 159} 
\usepackage{color}
\usepackage{ulem}

\journal{Elsevier}

\begin{document}

\title{MIMNet: Multi-Interest Meta Network with Multi-Granularity Target-Guided Attention for Cross-domain Recommendation }


\cortext[correspondingauthor]{Corresponding author} 

\author[1]{Xiaofei Zhu}
\ead{zxf@cqut.edu.cn} 

\author[1]{Yabo Yin}  
\ead{yinyabo@stu.cqut.edu.cn}

\author[2]{Li Wang\corref{correspondingauthor}}
\ead{wangli@tyut.edu.cn} 


\address[1]{College of Computer Science and Engineering, Chongqing University of Technology, Chongqing 400054, China}
\address[2]{College of Data Science, Taiyuan University of Technology, \\Shanxi Jinzhong 030600, China} 



\begin{abstract}
Cross-domain recommendation (CDR) plays a critical role in alleviating the sparsity and cold-start problem and substantially boosting the performance of recommender systems. Existing CDR methods prefer to either learn a common preference bridge shared by all users or  a personalized preference bridge tailored for each user to transfer user preference from the source domain to the target domain. Although these methods significantly improve the recommendation performance, there are still some limitations. First, these methods usually assume a user only has  a unique interest, while ignoring the fact that a user may interact with items with different interest preferences. Second,  they learn transformed preference representation mainly relies on the source domain signals, while neglecting the rich information available in the target domain. To handle these issues, in this paper, we propose a novel method named Multi-interest Meta Network with Multi-granularity Target-guided Attention (MIMNet) for cross-domain recommendation. To be specific, we employ  the capsule network to learn user multiple interests in the source domain, which will be fed into a meta network to generate multiple interest-level preference bridges. Then, we transfer user representations from the source domain to the target domain based on these multi-interest bridges. In addition, we introduce both fine-grained and coarse-grained target signals to aggregate user transformed interest-level representations by incorporating a novel multi-granularity target-guided attention network. 
We conduct extensive experimental results on three real-world CDR tasks, and the results show that our proposed approach MIMNet consistently outperforms all baseline methods. The source code  of MIMNet is released at https://github.com/marqu22/MIMNet.

\end{abstract}





\maketitle

\section{Introduction}

Recommender systems \cite{ zuo2023dynamic,zhu2024collaborative,guo2024lgmrec,wu2024personalized} have become indispensable in shaping the modern user experience in a wide range of domains, deeply influencing user choices in short videos \cite{yu2022mdp2,wang2023diffusion,bai2024labelcraft}, e-commerce \cite{xu2023multi,chen2024macro}, online recruitment \cite{hu2023boss,zheng2024mirror}, and etc.
Existing methods have been proven to be effective when rich user-item interaction information is available. However, in real application scenarios, some users would have few interaction information, especially for newly joined users (cold-start users),  which significantly hinders the application of a recommender system. 
To solve the problem, cross-domain recommendation (CDR) has been proposed, which attempts to alleviate the sparsity and cold-start issue in the target domain by exploiting knowledge from the source domain. 
Learning a common bridge to transfer user preference from the source domain to the target domain is a  general CDR strategy \cite{cite-EMCDR,cite-DCDCSR}. 
For example, EMCDR \cite{cite-EMCDR} employs a two-phase strategy to enhance the recommendation performance in the target domain. Specifically, it first encodes user embeddings in the source and target domains, respectively.  Then, a common bridge  between the source and target domains is learned based on overlapping users for aligning user embeddings in the two domains. 
Since the number of overlapping users would be inadequate for learning a high quality common bridge, some research works \cite{cite-SSCDR,cite-TMCDR,cite-PTUPCDR}  have been proposed to handle this issue. SSCDR \cite{cite-SSCDR} applies a semi-supervised framework to learn a common bridge between different domains and capture the neighboring information of users. 
TMCDR \cite{cite-TMCDR} employs the meta learning technique to learn a common bridge between the source and target domains. However, user preference usually varies from one to one, applying a common bridge to transfer user preference would lead to inferior performance. To the end, PTUPCDR \cite{cite-PTUPCDR} proposes to transfer user preference by using personalized bridges which take users' characteristic embeddings in the source domain as input. 

\begin{figure*}[!t]
    \centering 
\includegraphics[width=\textwidth]{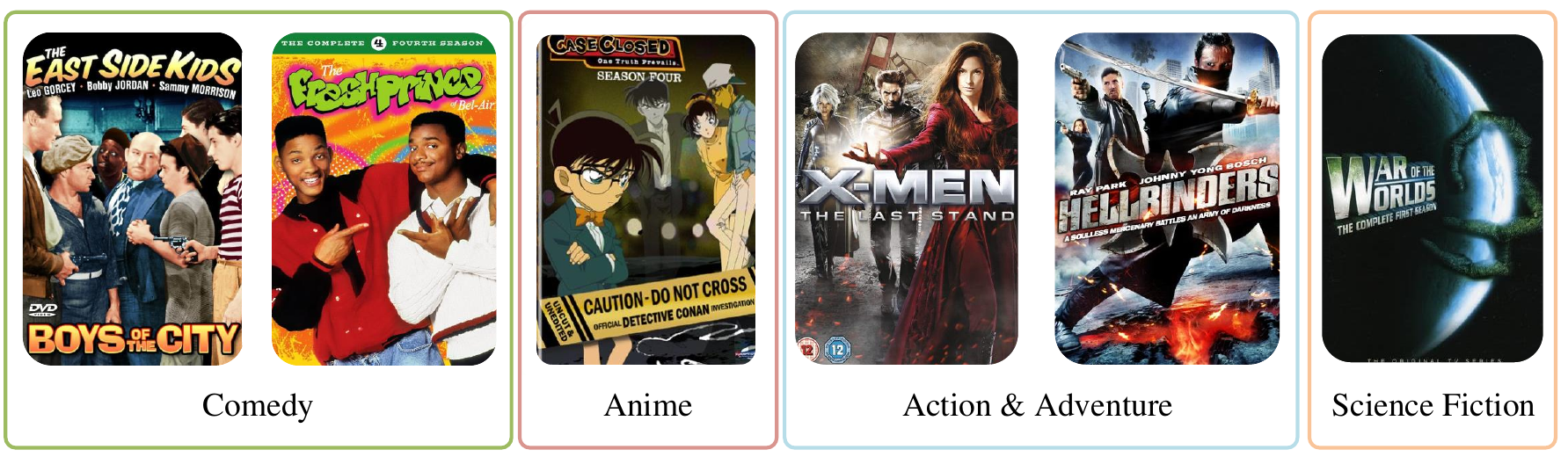}
\caption{An example of a user with different interest preferences on movie styles, including Comedy, Anime, Action \& Adventure and Science Fiction.}
\label{fig:cdr_case1}
\end{figure*}

Despite the significant successes achieved by these methods, they suffer from two limitations. First, these methods neglect the complexity of user interests where a user may have diverse interests rather than a unique interest. As depicted in Figure \ref{fig:cdr_case1}, a user has different interest preferences on movie styles, ranging from Comedy and Anime to Action \& Adventure and Science Fiction. However, the rich information of these interest-level preferences  are largely ignored by existing methods, as they either assume all users share a common preference transference or each user takes a specific preference transference.
Although there are a few efforts \cite{sun2023reinforced} trying to model users' multiple interests, they heavily rely on external knowledge such as item category and brand, which
may not available in real applications. 
Second, existing state-of-the-art methods learn transformed preference representation mainly based on information from the source domain. However, the rich information from the target domain is not well explored to guide the learning process.

To handle the above-mentioned issues, in this paper, we propose a multi-interest meta network with multi-granularity target-guided attention for cross-domain recommendation, termed MIMNet. To be specific, we first leverage the capsule network with dynamic routing to decouple user multiple interest representations. Then, the learned representations will be fed into a multi-interest meta network to generate a multi-interest preference bridge, which attempts to transfer user  preference from the source domain to the target domain in an interest-level manner. 
In addition, we   develop a multi-granularity target-guided attention network, which aims to incorporate fine-grained and coarse-grained target guidance to facilitate the aggregation of user representations with diverse interest preferences.
To guide the adaptive aggregation process of user preference in the target domain, the fine-grained guidance leverages the target item-level signal, while the coarse-grained guidance relies on the target prototype-level signal. 

We carry out extensive experiments on three real-world different CDR tasks to evaluate the effectiveness of our proposed approach. The results demonstrate that MIMNet is consistently superior to existing state-of-the-art baselines. The relative performance improvements of MIMNet  over  the best-performing baseline, REMIT \cite{sun2023reinforced}, on Task1 are 14.45\%, 16.36\%, 12.86\% when $\beta$ equals to 20\%, 50\%, 80\%  in terms of the MAE metric. Similar performance improvements can be observed on the other two Tasks. The main contributions of our work are summarized as:
\begin{itemize}
    \item We propose a novel multi-interest meta network  to  decouple users' multiple interests, and generating multi-interest bridges to transfer user embeddings from the source domain to the target domain.
    \item By exploring both fine-grained and coarse-grained target signals, we develop a multi-granularity target-guided attention network to adaptively guide the aggregation process of user representations with different interest preferences in the target domain. 
    \item We conduct extensive experiments on  three CDR tasks to validate the effectiveness of MIMNet. Experimental results show that our proposed  method outperforms state-of-the-art baseline methods.  
\end{itemize}


\section{RELATED WORK}

Cross-domain recommendation (CDR) attempts to alleviate the data sparsity and cold-start issue in the recommendation system by transferring user preference from the source domain to the target domain. It has substantially boosted the performance of recommendation and attracted the increasing attention of researchers. 
In early years, CMF \cite{cite-CMF} conducts cross-domain information transfer by applying matrix factorization across multiple domains to mitigate the problem of data sparsity in the target domain. 
CST \cite{cite-CST} utilizes coordinate systems for knowledge transfer between the source and target domains based on a principled matrix-based transfer learning framework.   
Recently, some researchers leverage deep learning-based CDR methods to model the collaborative relationships between different domains. 
CoNet \cite{cite-CoNet} assumes the hidden layers between two domains are connected, and proposes to apply dual knowledge transfer instead of one direction knowledge transfer.
MINDTL \cite{cite-MINDTL} facilitates  the modeling of user preference in the target domain by extracting and transferring rating patterns from the source domain. 
DDTCDR \cite{cite-DDTCDR} develops a latent orthogonal mapping to preserve  relations  between users among different domains, and transfers knowledge between these domains in an iterative way.

Other researchers attempt to learn a common bridge to transfer user preference between the two domains.
EMCDR \cite{cite-EMCDR}   models user preference in the source and target domains respectively, and then  learns a common bridge between the two domains based on the overlapping users. 
DCDCSR \cite{cite-DCDCSR} extends EMCDR by employing MF to learn user and item embeddings, and incorporates the fine-grained sparsity degrees of users and items to combine the learned embeddings. Due to the limited number of overlapping users, the two methods may result in unsatisfying performance. 
To handle this issue, SSCDR \cite{cite-SSCDR} employs a semi-supervised framework to learn a common bridge between different domains and capture the neighboring information of users. 
TMCDR \cite{cite-TMCDR} applies the meta learning technique to handle the above-mentioned issue based on its strong generalization ability. It develops a transfer-meta framework for CDR by learning a common bridge, which consists of two stages, i.e., a transfer stage and a meta stage. The former stage trains a source model and a target model on the source and target domains respectively, and the latter stage transforms user preference from the source domain to the target domain based on the common bridge.

Since the preference transition patterns of different users  between the source and target domains may vary considerably, PTUPCDR \cite{cite-PTUPCDR}  utilizes a personalized bridge  rather than a single common bridge to transfer each user's specific preference. It generates  a personalized bridge for each user by taking their characteristic representations in the source domain as input of the bridge.
REMIT \cite{sun2023reinforced} proposes to extract users' multiple interests in the source domain based on different meta-path based aggregations, and utilizes a reinforcement learning framework to aggregate transformed interests. 
The main difference between our method and REMIT are three aspects. 
First, REMIT relies on external knowledge such as item category and brand, which may not available in real applications, to construct different meta-paths. In contrast, our method only utilizes user interaction data which is more efficient and practical. 
Second,  REMIT learns transformed preference representation by mainly leveraging the information in the source domain while neglecting the rich information from the target domain.  
However, our method proposes to introduce informative target signal to guide the aggregation of user transformed representations. 
Third, REMIT transfers user preference based on  a single granularity which overlooks the inherent multiple granularity property of user preference. Different from REMIT, our method  develops both fine-grained and coarse-grained target guidance to facilitate the
multi-granularity transformation of user diverse preferences.

\section{Problem Formulation}
In the task of cross-domain recommendation (CDR), there are two domains, i.e., a source domain $\mathcal{D}_s$ and a target domain $\mathcal{D}_t$. In each domain, we have a user set $\mathcal{U}=\{u_1, u_2, \cdots, u_m\}$,  an item set $\mathcal{V}=\{v_1, v_2, \cdots, v_n\}$  and a rating matrix $\mathcal{R}$, where $r_{ij} \in \mathcal{R}$ indicates  there is an interaction between user $u_i$ and item $v_j$, $m$ and $n$ are the number of users and items, respectively.
We use $\mathcal{U}^s$, $\mathcal{V}^s$, and $\mathcal{R}^s$ to denote the user set, the item set, and the rating matrix in the source domain. Similarly, we use $\mathcal{U}^t$, $\mathcal{V}^t$ and $\mathcal{R}^t$ for the target domain. 
The overlapping users between the source and target domains are defined as $\mathcal{U}^o = \mathcal{U}^s \cap \mathcal{U}^t$. It is worth noting that there are no shared items between the two domains, i.e., $\mathcal{V}^s \cap \mathcal{V}^t = \emptyset$. The cold-start users denoted by $\mathcal{U}^c=\{u|u\in \mathcal{U}^s \land u \notin \mathcal{U}^t\}$ are those who have interactions with items in the source domain while having no interactions with items in the target domain. 
For each user $u_{i}^{s} \in \mathcal{U}^s$ in the source domain, we denote ${S}_{u_i}=\{v_{1}^s,v_{2}^s 
\ldots,v_{n_{i}}^s\}$  as her corresponding  interacted items, where $n_i$ and $v_{j}^s$ are the number of interacted items and the $j$-th  interacted item  of $u_{i}^{s}$. 
We can transform the users and items into dense vectors, also called embeddings, with the latent factor model \cite{jenatton2012latent}. In this paper, we use $\textbf{u}_i^{*}\in\mathbb{R}^{d}$  and $\textbf{v}_j^{*}\in\mathbb{R}^{d}$ to denote the embeddings of the user $u_i^*$ and the item $v_j^{*}$ respectively, where $d$ denotes the dimensionality of the embedding and $*\in\{s,t\}$ represents the label of domain. Specifically, we use  $\textbf{S}_{u_{i}}=\{\textbf{v}_{1}^s,\textbf{v}_{2}^s,\ldots,\textbf{v}_{n_{i}}^s\}$ to denote the embeddings of interacted items of  $u_{i}^{s}$ in the source domain, where  $\textbf{v}_j^s\in\mathbb{R}^{d}$ is the  embedding of the interacted item $v_j^s$.   
The goal of CDR is to improve the performance of recommendations in the target domain by leveraging rich information from the source domain.

\section{Our Proposed Model} 
\begin{figure*}[!th]
    \centering 
\includegraphics[width=\textwidth]{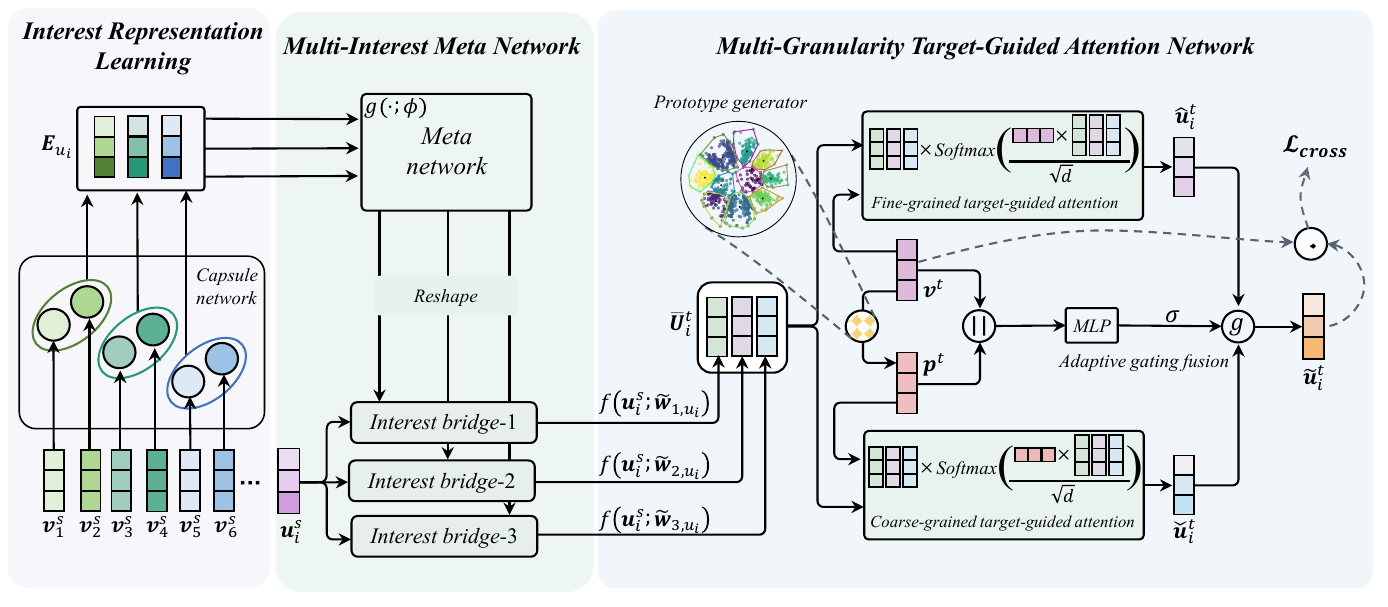}
\caption{Overall architecture of our proposed MIMNet model, which consists of three components: (1) interest representation learning, (2) multi-interest meta network, (3) multi-granularity target-guided attention network. }
\label{fig:framework}
\end{figure*}

The architecture of our proposed MIMNet model is illustrated in Figure \ref{fig:framework}, which contains three components, i.e., interest representation learning,  multi-interest meta network, and multi-granularity target-guided attention network.

\subsection{Interest Representation Learning} 
In this sub-section, we attempt to learn the multiple interest representations of a user based on her sequential interaction items.
Inspired by \citep{sabour2017dynamic,li2019multi}, we utilize dynamic routing in capsule network to  extract users' multiple interests. To make this paper self-contained, we briefly revisit the key basics of dynamic routing.
In dynamic routing, there are two layers of capsules, including  low-level capsules and high-level capsules. The two layers of capsules are updated in an iterative way, and the final high-level capsules are considered as user extracted interests.

Given the embeddings of sequential interaction items for a user $u$ in source domain, i.e., $\textbf{S}_{u} = \{\textbf{v}^s_1,\textbf{v}^s_2,\cdots,\textbf{v}^s_n\}$. Our goal is to extract $K$ interests $\textbf{e}_k \in \mathbb{R}^{h}$, $k \in \{1,\cdots,K\}$.
In each iterative process of dynamic routing, we first compute the routing logit score $b_{jk}$ as:
\begin{equation}
   b_{jk} ={{(\textbf{e}_{k}})^T}\textbf{M}\, \textbf{v}^s_{j},
\end{equation}
where the $\textbf{M} \in \mathbb{R}^{h \times d}$ denotes the transformation matrix to be learned.
With the routing logit scores calculated, the candidate vector  $\textbf{z}_{k}$ for the $k$-th high-level capsule  is computed as a weighted sum of all low-level capsules. The calculation of $\textbf{z}_{k}$ is as follows:
\begin{equation}
   \textbf{z}_{k} = {\sum_{j=1}^n}w_{jk}\textbf{M}\,\textbf{v}^s_j,
\end{equation}
\begin{equation}
   w_{jk} = \frac{exp(b_{jk})}{\textstyle \sum_{m=1}^K{exp(b_{jm})}},
\end{equation}
where $w_{jk}$ denotes the contribution weight score of the $j$-th low-level capsule to the ${k}$-th high-level capsule. 

Then, a non-linear ``squash'' function is used to ensure high-level capsule vectors are in an appropriate range:
\begin{equation}
   \mathbf{e}_{k}=\operatorname{squash}(\mathbf{z}_{k})=\frac{\left\|\textbf{z}_{k}\right\|^{2}}{1+\left\|\textbf{z}_{k}\right\|^{2}} \frac{\textbf{z}_{k}}{\left\|\textbf{z}_{k}\right\|},
\end{equation}
where $\textbf{e}_k$ is the $k$-th interest we want to extract from the sequential interaction items in the source domain. For each user ${u_i}$, $K$ interest vectors can be extracted from the interaction sequence in the source domain through the dynamic routing method. We combine these $K$ interest vectors into a matrix $\textbf{E}_{u_i} = [\textbf{e}_1,\cdots ,\textbf{e}_K] \in \mathbb{R}^{K \times d}$ to serve the downstream task. 

\subsection{Multi-Interest Meta Network}

The core of cross-domain recommendation task is to transfer user  preference from the source domain to the target domain. To the end, existing methods either learn a common preference bridge shared by all users \cite{zhu2021transfer} or leverage a personalized preference bridge tailored for each user \cite{cite-PTUPCDR}. Although these methods have achieved promising performance, they neglect the fact that a user may interact with items with multiple interest preferences. To solve this issue, we resort to leverage interest-level preference bridges. Specifically, we first employ a meta network that takes the multiple interests $\textbf{E}_{u_i}$  of a user in the source domain as input to generate multi-interest bridges, and then we transfer her embeddings from the source domain  to the target domain based on the multi-interest bridges. 
Formally, we have:
\begin{equation}
    \textbf{W}_{u_i} = g(\textbf{E}_{u_i} ; \phi),
\end{equation}
where $g(\cdot)$ is a two layer feed-forward neural network, $\phi$ are learnable parameters, and $\textbf{W}_{u_i} = [ \textbf{w}_{1,u_i},\cdots,\textbf{w}_{K,u_i} ]\in \mathbb{R}^{K\times d^2}$  with $\textbf{w}_{k,u_i}\in \mathbb{R}^{d^2}$ be the generated parameter vector for the $k$-th interest bridge. 
We reshape $\textbf{w}_{k,u_i}$ into a matrix $\tilde{\textbf{w}}_{k,u_i} \in \mathbb{R}^{d \times d}$, and formulate the $k$-th interest bridge for user $u_i$ as:
\begin{equation}
     f(\cdot ; \tilde{\textbf{w}}_{k,u_i}).
\end{equation}
, where $f(\cdot)$ is a linear layer. With the  learned multi-interest bridges for $u_i$, we can obtain her corresponding transformed representations in the target domain. Formally,  we have:
\begin{equation}
     \bar {\textbf{u}}^t_{k,i} = f(\textbf{u}^s_i ; \textbf{w}_{k,u_i}),
\end{equation}
where $\textbf{u}^s_i \in \mathbb{R}^d$ denotes the embedding of user $u_i$ in the source domain, and $\bar{\textbf{u}}^t_{k,i}  \in \mathbb{R}^d$ denotes the transformed embedding of user $u_i$ in the target domain based on the $k$-th interest bridge. As there are $K$ transformed embeddings for $u_i$, i.e., $\{\bar{\textbf{u}}^t_{k,i}\}_{k=1}^K$, we combine these transformed embeddings into a matrix $\bar{\textbf{U}}^t_{i} =[\bar {\textbf{u}}^t_{1,i},...,\bar {\textbf{u}}^t_{K,i}]\in \mathbb{R}^{d \times K}$, which will be utilized to obtain the final representation for $u_i$ in the target domain.

\subsection{Multi-Granularity Target-Guided Attention Network}
Since the transformed embeddings $\bar{\textbf{U}}^t_{i}$ of $u_i$  solely rely on information from the perspective of source domain, while the informative signals from the target domain are ignored. 
As information from the target domain also plays a critical role in guiding the transfer process, we further propose to take into account the target signal to learn better transformed user representation in the target domain. Specifically, we develop a novel multi-granularity target-guided attention network which leverages both fine-grained and coarse-grained target signals to aggregate a user's different interest preferences.

\subsubsection{Fine-grained Target-Guided Attention}

In this module, we attempt to aggregate the learned multiple interest preferences $\bar{U}^t_{i}$ into a transformed user embedding $\hat{u}^t_{i}$ with the guidance of fine-grained target signals, i.e., the information from each candidate item $v^t_j$ in the target domain. Let $\textbf{v}^t_j$ be the corresponding embedding of $v^t_j$. We employ attention mechanism to aggregate the multiple interest preferences  $\bar{U}^t_{i}$, where we take $\textbf{v}^t_j$ as the query, and $\bar{U}^t_{i}$ as both keys and values. The fine-grained target-guided aggregation is defined as follows:
\begin{equation}
\begin{split}
\hat{\textbf{u}}^t_i &= Attention(\textbf{v}^t_j, \bar{\textbf{U}}^t_i,\bar{\textbf{U}}^t_i)   \\
\end{split}
\label{eq:tgt-item-attention}
\end{equation} 
where $\hat{\textbf{u}}^t_i \in \mathbb{R}^d$ is the user's transformed embedding in the target domain from a fine-grained perspective.

\subsubsection{Coarse-grained Target-Guided Attention}
 
To improve the generalization ability of our model, we propose to extract the prototype of each candidate item in the target domain and leverage it a coarse  signal to aggregate the learned multiple interest preferences $\bar{U}^t_i$.  To be specific, we employ K-Means\footnote{https://github.com/facebookresearch/faiss} to cluster items in the target domain  and utilize the center of each cluster as the  prototype of  items in the cluster. Denote $\textbf{p}^t_j \in \mathbb{R}^d $ as the corresponding prototype of $v^t_j$, and we define the process of the coarse-grained target-guided aggregation as follows: 
\begin{equation}
\check{u}^t_i = Attention({p}^t_j, \bar{U}^t_i,\bar{U}^t_i),
\end{equation} 
where $\check{u}^t_i \in \mathbb{R}^d$ denotes the user's embeddings generated by the prototype of the candidate item guide.

After we obtain $\hat{u}^t_i$ and $\check{u}^t_i$, we introduce an adaptive fusion module to generate the final  user's embeddings $\tilde{u}^t_i$ in the target domain, which is defined as:
\begin{align}
\tilde {\textbf{u}}^t_i &= \alpha  \cdot \hat{\textbf{u}}^t_i + (1-\alpha)  \cdot \check{\textbf{u}}^t_i,\\
\alpha &= \sigma \left(MLP\left([\textbf{v}^t_j ; \textbf{p}^t_j]\right)\right),
\end{align} 
where $[;]$ denotes the concatenation operator and $\sigma$ is a $sigmoid(\cdot)$ function.

\subsection{Prediction and Model Optimization}
\subsubsection{Prediction.} Given a candidate item $v_j$, we first utilize the target-guided adaptive fusion module to derive the user's embedding $\tilde {u}^t$.
Then, we apply the inner product to calculate the user $u_i$ ratings for the candidate item $v_j$ as follows:
\begin{equation}
    \hat{y}_{ij} = (\tilde {u}^t_i)^T  v_j,
\end{equation}
where $\hat{y}_{ij}$ denotes predicted rating of user $u_i$ for candidate item $v_j$.
\subsubsection{Model Optimization.} 
The overall model consists of two stages of training, including a pre-training stage and a cross-domain training stage.

\textbf{Pre-training stage: }   
The goal of the pre-training step is to learn embeddings of   users and items in the source domain and target domain, respectively. As interactions of the overlapping users in the target domain are not visible in this stage, we use 
The loss function in the pre-training stage is formulated as:
\begin{equation}
    \min_{u,v} \frac{1}{|\mathcal{ R} |} \sum _{r_{ij} \mathcal{ \in  R}}(r_{ij}-u_{i}^T v_{j})^2 .
\end{equation}

\textbf{Cross-domain training stage: } 
In the cross-domain recommendation stage, we employ the task-oriented training strategy, which utilizes the final rates  as the optimization goal rather than leverages user embeddings in the target domain for optimization.
Specifically, the loss function in the cross-domain training stage is formulated as:
\begin{equation}
 \min _{} \frac{1}{|\mathcal{R}_{o}^{t}|} \sum_{r_{i j} \in \mathcal{R}_{o}^{t}}(r_{i j}- \hat{y}_{ij})^{2}.
\end{equation}


\section{EXPERIMENTS}

In this section, we first conduct extensive experiments on three cross-domain tasks under different cold-start settings to evaluate the performance of our proposed method.
Then, we analyze the effect of using different base models for our method and investigate the effectiveness of the different numbers of interests on our model performance.
Finally, we analyze  the magnitude of the contribution of different modules in the model to the overall effect of the model by using ablation experiments and visualized the analysis.
\subsection{Experimental Settings}


\begin{table*}[hbt!]
\begin{center}

    \caption{The data statistics and task definitions, where Overlap denotes overlapping users in both domains.}
    \begingroup
    \resizebox{0.95\linewidth}{!}{  
    \label{tab:dataset}
    \tabcolsep=0.12cm  
    \setlength{\tabcolsep}{6pt} 
    \renewcommand{\arraystretch}{1} 
    \begin{tabular}{ c | c c | c c | c c c | c c}
    
    \toprule
    \multirow{2}{*}[-0.75ex]{{CDR Tasks}} & 
    \multicolumn{2}{c|}{{Domain}} & 
    \multicolumn{2}{c|}{{Item}} & 
    \multicolumn{3}{c|}{{User} } & 
    \multicolumn{2}{c }{{Rating}}  \\ 
    \cmidrule(){2-10}
     ~ & Source & Target & Source & Target &  Overlap & Source & Target & Source & Target \\ 
    
    \midrule
    {Task1} & Movie & Music & 50,052 & 64,443 & 18,031 & 123,960 &  75,258 & 1,697,533 &  1,097,592\\ 
    
    {Task2} & Book & Movie & 367,982 &  50,052 & 37,388 &  603,668 &  123,960 & 8,898,041 &  1,697,533\\ 
   
    {Task3} & Book & Music & 367,982 &  64,443 & 16,738 &  603,668 & 75,258 & 8,898,041 & 1,097,592\\ 
    
    \bottomrule
    \end{tabular}
    }
    \endgroup
\end{center}    
\end{table*}

\subsubsection{{Datasets}} 
Amazon review dataset\footnote{http://jmcauley.ucsd.edu/data/amazon/} is a real-world public dataset which has been widely used in the task of CDR \cite{cite-SSCDR,zhu2021transfer,cite-PTUPCDR}.
Following most existing works, we utilize the Amazon dataset, in which users rated items in a range of 1-5.
Specifically, the dataset has 24 categories, in which 3 popular categories, i.e., movies\_and\_tv (Movie), cds\_and\_vinyl (Music), and books (Book), are selected for the experiments. Table \ref{tab:dataset} shows the detailed statistics of the dataset.

\subsubsection{{Baselines}} 
To evaluate the performance of our proposed approach, we choose the following baselines for comparison.
\begin{itemize} 
    \item{\textbf{TGT}} \cite{cite-PTUPCDR}. TGT is a method which is trained by only leveraging information from the target domain.
    \item{\textbf{CMF}} \cite{cite-CMF}. CMF is an extended version of MF, which enables cross-domain recommendation for users by sharing an embeddings for the same user in the source domain and target domain.
    \item{\textbf{EMCDR}} \cite{cite-EMCDR}. EMCDR is a widely used cross-domain recommendation method. It applies a common bridge to transfer user preference from the source domain to the target domain. Specifically, it first utilizes  matrix factorization (MF) to learn user and item embeddings on each domain, and then adopts a common bridge to capture the underlying relationship between the two domains.
    \item{\textbf{DCDCSR}} \cite{cite-DCDCSR}. Similar to EMCDR, this method also employs MF to learn user and item embeddings. However, it combines the learned embeddings by introducing the fine-grained sparsity degrees of users and items which can utilize more rating data in both domains.
    \item{\textbf{SSCDR}} \cite{cite-SSCDR}. It first models user-user similarities and user-item interactions, and then learns a common bridge between different domains with a semi-supervised framework. To infer the latent representations of the cold-start users, it further takes into account the neighbors of these users.
    \item{\textbf{PTUPCDR}} \cite{cite-PTUPCDR}. Compared with previous works, PUTPCDR attempts to learn a personalized bridge rather than a common bridge  to transfer user preference.  To generate the personalized bridge, it learns a meta network based on users' characteristics in the source domain.
    \item{\textbf{REMIT}} \cite{sun2023reinforced}. This method employs a heterogeneous information network and different meta-path based aggregations to extract user interests in the source domain. 
    A reinforcement learning-based strategy is then used to transfer and aggregate users' interests to the target domain. Different from existing methods, REMIT relies on additional knowledge such as categories and brands of items.

\end{itemize}
\subsubsection{{Evaluation Metrics}} In the experiments, we adopt two metrics: Mean absolute Error (MAE) and Root Square Error (RMSE) to measure the performance of all methods following  \cite{cite-EMCDR,cite-PTUPCDR}, which are widely used in the task of CDR.

\subsubsection{{Implementation Details}} 
We set the learning rate for the Adam optimizer to 0.01, the size of mini-batch to 512, and  the embedding dimension $d$ to 10. The iteration number of dynamic route is fixed to 3, and the number of prototypes in the target domain is set to 100. 
To evaluate the performance of our proposed method, we randomly select some overlapping users and remove their ratings in the target domain, and regard these users as test users. The remaining overlapping users are utilized for training multi-interest bridges.  
Similar to \cite{cite-PTUPCDR,sun2023reinforced}, we set the ratio of test users $\beta$ to 20\%, 50\%, 80\% of all overlapping users, respectively. 
For each task, the averaged results over five random runs are reported.
\subsection{Performance Comparison}

\begin{table*}[!ht]
\begin{center}
    \caption{The overall performance in different cold-start scenarios for three CDR tasks. The best and second best scores are in bold and underlined, respectively. 
    Results marked with {$\star$} and $\ast$  are taken from \cite{cite-PTUPCDR} and  \cite{sun2023reinforced}, respectively. We report the reimplemented results of EMCDR and PTUPCDR for comparision as they are better than the ones reported in \cite{cite-PTUPCDR}. (Note that a lower MAE and RMSE value indicates a better model performance) }
    \begingroup
    \resizebox{1\linewidth}{!}{  
    \label{tab:overall}
    \setlength{\tabcolsep}{2pt} 
    \begin{tabular}{ c|c | c | c c @{\hspace{3pt}} c @{\hspace{3pt}} c @{\hspace{6pt}} c @{\hspace{6pt}} c c | c c }
    \toprule
    {\textbf{~}} & 
    {\textbf{$\beta$}} & 
    {{Metric}} & 
    {\enspace{TGT}\textsuperscript{$\star$} } & 
    {\enspace{CMF}\textsuperscript{$\star$}}  &
    {\enspace{DCDCSR}\textsuperscript{$\star$}} & 
    {\enspace{SSCDR}\textsuperscript{$\star$}} & 
    {{EMCDR}} & 
    {{PTUPCDR}} & 
    {\enspace{REMIT}\textsuperscript{$\ast$}}  & 
    {{MIMNet}} &
    {{Improv.}}  \\

    \midrule
    
     \multirow{6}{*}[-1ex]{{Task1}} & \multirow{2}{*}{{20\%}} & MAE  & 4.4803 &  1.5209 & 1.4918 & 1.3017  & 1.1141  & 1.0478 & \underline{0.9393} & \textbf{0.8027} & 14.45\%\\ 
     ~                                     &   ~                            & RMSE & 5.1580 &  2.0158 & 1.9210 & 1.6579  & 1.3881  & 1.3693 & \underline{1.2709} & \textbf{1.1509} & 9.44\%\\ 
     \cmidrule(){2-12}
     
     ~                                     & \multirow{2}{*}{{50\%}} & MAE  & 4.4989 & 1.6893 & 1.8144 & 1.3762   & 1.2780   & 1.1340  & \underline{1.0437}  & \textbf{0.8729} &16.36\%\\
     ~                                     &  ~                             & RMSE & 5.1736 & 2.2271 & 2.3439 & 1.7477   & 1.5738   & 1.4982  & \underline{1.4580}  & \textbf{1.2244} &16.02\%\\
     \cmidrule(){2-12}
     ~                                     & \multirow{2}{*}{{80\%}} & MAE  & 4.5020 & 2.4186 & 2.7194 & 1.5046   & 1.7345    & 1.3786 & \underline{1.2181} & \textbf{1.0614}   &12.86\%\\
     ~                                     &  ~                             & RMSE & 5.1891 & 3.0936 & 3.3065 & 1.9229   & 2.0977   & 1.8883 & \underline{1.6601} & \textbf{1.4721}   &11.32\%\\
     \midrule
     \multirow{6}{*}[-1ex]{{Task2}} & \multirow{2}{*}{{20\%}} & MAE  & 4.1831 & 1.3632 & 1.3971 & 1.2390   & 0.9492   & 1.0093 & \underline{0.8759} & \textbf{0.8718}  &00.47\%\\ 
     ~                                     &   ~                            & RMSE & 4.7536 & 1.7918 & 1.7346 & 1.6526   & 1.1887   & 1.2947 & \underline{1.1650} & \textbf{1.1430}  &01.89\%\\
     \cmidrule(){2-12}
     ~                                     & \multirow{2}{*}{{50\%}} & MAE  & 4.2288 & 1.5813 & 1.6731 & 1.2137   & 1.0064   & 1.0428 & \underline{0.9172} & \textbf{0.9025}  &01.60\%\\
     ~                                     &  ~                             & RMSE & 4.7920 & 2.0886 & 2.0551 & 1.5602   & 1.2558   & 1.3519 & \underline{1.2379} & \textbf{1.1983}  &03.20\%\\
     \cmidrule(){2-12}
     ~                                     & \multirow{2}{*}{{80\%}} & MAE  & 4.2123 & 2.1577 & 2.3618 & 1.3172   & 1.1330   & 1.1149 & \underline{1.0055} & \textbf{0.9710}   &03.43\%\\
     ~                                     &  ~                             & RMSE & 4.8149 & 2.6777 & 2.7702 & 1.7024   & 1.4388   & 1.4756 & \underline{1.3772} & \textbf{1.2910}   &06.26\%\\
     \midrule
     \multirow{6}{*}[-1ex]{{Task3}} & \multirow{2}{*}{{20\%}} & MAE  & 4.4873 & 1.8284 & 1.8411 & 1.5414   & 1.3302   & \underline{1.1241}  & 1.3749 & \textbf{0.8107}   &27.88\%\\ 
     ~                                     &   ~                            & RMSE & 5.1672 & 2.3829 & 2.2955 & 1.9283   & 1.5923   & \underline{1.4728}  & 1.9940 & \textbf{1.1711}   &20.48\% \\ 
    \cmidrule(){2-12}
     ~                                     & \multirow{2}{*}{{50\%}} & MAE  & 4.5073 & 2.1282 & 2.1736 & 1.4739   & 1.6004   & \underline{1.2566}  & 1.4401 & \textbf{0.9348}   &25.61\%\\
     ~                                     &  ~                             & RMSE & 5.1727 & 2.7275 & 2.6771 & 1.8441   & 1.9130   & \underline{1.6939}  & 2.0495 & \textbf{1.3009}   &23.20\% \\
     \cmidrule(){2-12}
     ~                                     & \multirow{2}{*}{{80\%}} & MAE  & 4.5204 & 3.0130 & 3.1405 & 1.6414   & 1.9968   & \underline{1.5122} & 1.6396 & \textbf{1.1167}   &26.15\%\\
     ~                                     &  ~                             & RMSE & 5.2308 & 3.6948 & 3.5842 & 2.1403   & 2.3634   & \underline{2.0825} & 2.2653 & \textbf{1.5178}   &27.12\%\\
    
    \bottomrule
    \end{tabular}
    }
    \endgroup
\end{center}   

\end{table*}
\subsubsection{{Overall}} In this section, we analyze the performance of our proposed MIMNet method  in different cold-start scenarios, and  the overall results of all comparing methods are shown in Table \ref{tab:overall}. We have the following key observations:
\begin{itemize}
\item The performance of TGT  is the worst among all baseline methods. This is because  TGT solely relies on information from the target domain to conduct the recommendation while ignoring the rich information from the source domain. 
CMF demonstrates a superior performance as compared with TGT since it can leverage information from both source and target domains.  
\item Although CMF achieves a better performance, it simply combines information of different domains into a single domain which inevitably ignores the potential domain drift.
DCDCSR, SSCDR, and EMCDR outperform CMF as they further capture the potential domain drift by introducing a common bridge to transfer user embeddings from the source domain into the target domain. 
\item By comparing PTUPCDR with these common bridge based methods, we can observe a considerable performance improvement brought by learning personalized bridges for each user.
Among baseline methods, REMIT obtains the best performance since it employs multiple personalized bridges together with a RL-based bridge sector to select transformed interests. 
\item Our proposed method MIMNet yields the best performance  on all three tasks under all different cold-start settings. For example, the relative performance improvements with $\beta=20\%$ of MIMNet over the best performing baseline method REMIT on Task1  in terms of MAE and RMSE are 14.45\% and 9.44\%, respectively. 
The main reason are that MIMNet can  extract better user multiple interests based on the capsule network and the meta network. In addition, MIMNet incorporates a multi-granularity target-guided  aggregation module to effectively aggregate the learned multiple interests.
\end{itemize}

\begin{figure*}[!t]
    \centering
    \includegraphics[width=\textwidth]{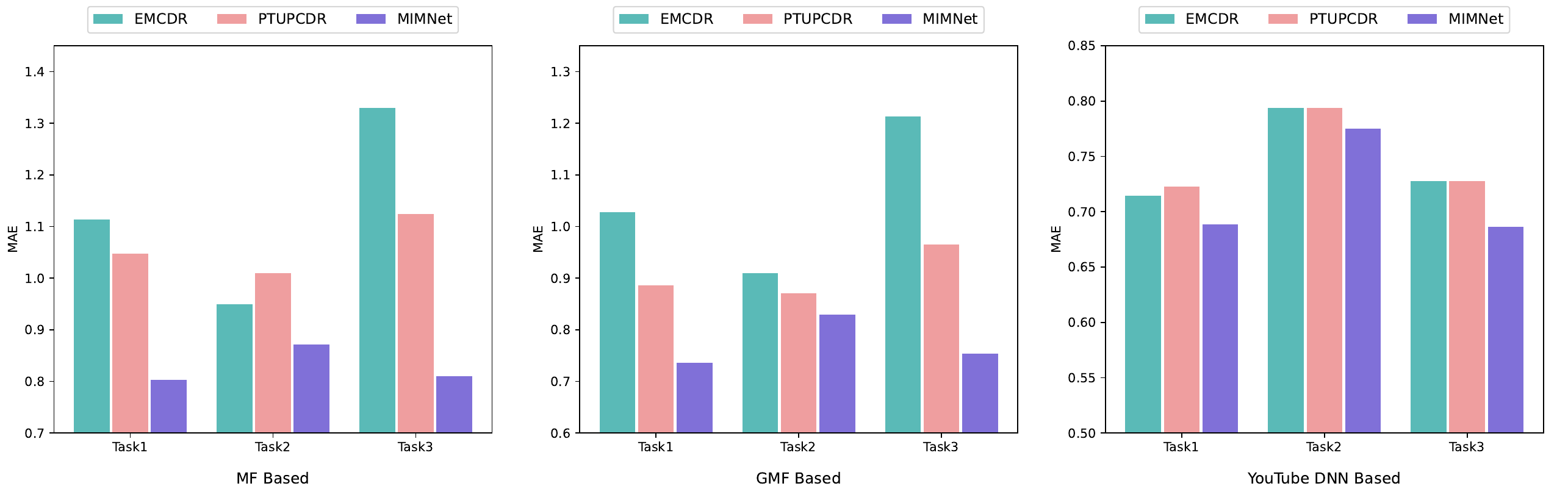}
    \caption{Performance comparison by applying EMCDR, PTUPCDR, and  MIMNet upon three base models MF, GMF and YouTube DNN with $\beta =20\%$, in terms of MAE.}
    \label{fig:different_base}
\end{figure*}
\subsubsection{{Generalization Experiments} }
%
Most of the previous  CDR methods \cite{cite-EMCDR,cite-DCDCSR} have predominantly focused on the design of the bridge for cross-domain mapping, utilizing relatively simple non-neural model (i.e., MF) as the base models.
However, considering the diversity of recommendation models, this simplification raises concerns about the  generalization capability of bridge-based CDR methods.
Following \cite{cite-PTUPCDR}, we introduce two additional neural network models, GMF \cite{he2017neural} and YouTube DNN \cite{covington2016deep}, as foundational domain models to validate the effectiveness of our proposed MIMNet method.
We compare MIMNet with two bridge-based approaches, i.e., EMCDR and PTUPCDR, upon all the three base models. 
It is worth noting that we neglect the baseline REMIT as it relies on additional information such as item categories and brand names.
The experimental results are shown in Figure \ref{fig:different_base} , and we can have the following observations: 
\begin{itemize}
\item  By replacing the base model MF with GMF, all three methods have shown considerable performance improvements. In addition, when we  employ  YouTube DNN as the base model, it will further boost the performance of all three methods. 
\item PTUPCDR generally outperforms EMCDR across the three different base models, indicating the superiority of using the personalized bridge over the  common bridge to transfer user preference.

\item Additionally, it can be observed that our proposed method MIMNet consistently exhibits the best performance across all base models, which reflects the generalization capability of MIMNet.
\end{itemize}

 \subsection{Model Analysis}
\begin{table}
\begin{center}
    \caption{Ablation study of MIMNet, in terms of MAE. }
    \label{tab:ablation_study1}
    \begingroup
    \tabcolsep=0.12cm  
    \renewcommand{\arraystretch}{1} 
    \resizebox{0.85\linewidth}{!}{  
    \begin{tabular}{ c | c c c | c c c | c c c}
    \toprule
    \multirow{2}{*}[-0.75ex]{Methods} & 
    \multicolumn{3}{c|}{Task1} & 
    \multicolumn{3}{c|}{Task2} & 
    \multicolumn{3}{c}{Task3}  \\ 
    \cmidrule(){2-10}
       ~                 & $20\%$           & $50\%$           & $80\%$           & $20\%$           & $50\%$           & $80\%$  & $20\%$           & $50\%$           & $80\%$  \\
    \midrule
    MIMNet                  & \textbf{0.8027} & \textbf{0.8729}  & \textbf{1.0614} 
                            & \textbf{0.8718} & \textbf{0.9025}  & \textbf{0.9710} 
                            & \textbf{0.8107} & \textbf{0.9348}  & \textbf{1.1167}\\
    \midrule
    -w/o multi\;            &        {0.8402} &        {0.9773}  &        {1.3399} 
                            &        {0.8852} &        {0.9145}  &        {1.0298} 
                            &        {0.8844} &        {1.1490}  &        {1.4744} \\
                            
-w/o target \hspace{-0.3em} &        {0.9828} &        {1.0977}  &        {1.3956} 
                            &        {0.9139} &        {0.9517}  &        {1.0515} 
                            &        {0.9995} &        {1.1820}  &        {1.4770} \\

    -w/o proto\;            &        {0.9531} &        {1.0471}  &        {1.2648} 
                            &        {0.9107} &        {0.9448}  &        {1.0231} 
                            &        {0.9648} &        {1.0919}  &        {1.2987} \\

    -w/o  adapt             &        {0.9378} &        {1.0170}  &        {1.1985} 
                            &        {0.9080} &        {0.9398}  &        {1.0049} 
                            &        {0.9455} &        {1.0503}  &        {1.2283} \\

    \bottomrule
    \end{tabular}
    }
    \endgroup
\end{center}
\end{table}

\subsubsection{{Ablation Study} }
To verify the effectiveness of the main components in MIMNet, we conduct an  additional ablation study in this section. Specifically, we consider the following variants of MIMNet for experiments:
\begin{itemize}
    \item \textbf{\textit{MIMNet w/o multi}}: we utilize a single-interest meta network instead of a multi-interest meta network to  transfer user preference.
    \item \textbf{\textit{MIMNet w/o target}}: we remove the fine-grained target-guided aggregation module, where   
    the fine-grained signal based on each candidate items will be ignored.  
    \item \textbf{\textit{MIMNet w/o proto}}:  we remove the  coarse-grained target-guided aggregation module, where the coarse-grained signal based on the prototypes of each candidate item will be overlooked.
    \item \textbf{\textit{MIMNet w/o adapt}}: we replace the adaptive gating fusion module with a mean-pooling.    
\end{itemize}

As shown in Table \ref{tab:ablation_study1}, our proposed method MIMNet   obtains the best performance  compared to other variants. 
Specifically, replacing the multiple interests with only a single interest leads to a significant performance degradation. 
This experimental result illustrates that by decoupling multiple interests within user interaction sequences in source domain, a more enriched representation of user preference can be obtained compared to modeling methods based on a single interest.
Additionally, the fine-grained signal based on each candidate item in the target domain is beneficial for guiding the aggregation process of multiple transformed interest representations. Moreover, discarding the  coarse-grained target-guided aggregation module results in a considerable performance decay, indicating the necessity to introducing coarse-grained guided signals from the target domain.
At last, adaptively fusing the fine-grained and coarse-grained transformed embeddings of a user will obtain better performance  as compared to  the strategy of applying the mean-pooling.

\subsubsection{{Performance with Different Interaction Sparsity Degrees} }
\begin{figure*}[!t]
    \centering
    \includegraphics[width=\textwidth]{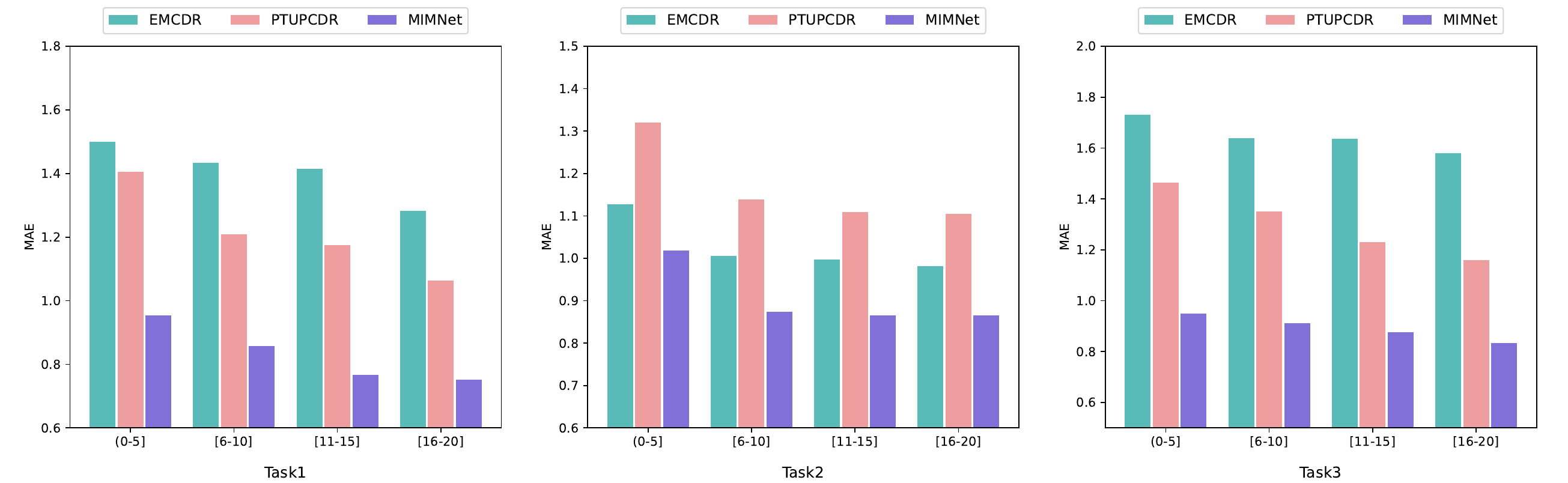}
    \caption{ Impact of different interaction sparsity degrees in the source domain on model performance with $\beta =20\%$, in terms of MAE.}
    \label{fig:buck}
\end{figure*}
To investigate the performance of our proposed model under different interaction sparsity degrees, we group users in the source domain into four categories, i.e., (0,5], [6,10], [11,15], [16,20]. Figure \ref{fig:buck} illustrates the performance comparison between our proposed method (i.e., MIMNet) and two competitive baseline methods (i.e., EMCDR and PTUPCDR).   
We observe that our proposed method MIMNet consistently outperforms  the two baseline methods across all interaction sparsity degrees, which reveals the robustness of our model under different interaction sparsity degrees. 
%
%
In addition, the two baselines have their own performance advantages in different tasks, e.g., EMCDR outperforms PTUPCDR in Task1 and Task3, while it is inferior to PTUPCDR in Task2. This result is attributed to that Task2 have relatively larger number of overlapping users as compared to the other two tasks. 
In contrast, our model performs better than both EMCDR and PTUPCDR in all tasks, which indicates the superiority of MIMNet among different task scenarios.
%
%
Moreover, the results also show that the performance of our method improves gradually when the number of user interactions increases. This reflects that our model can effectively capture user preference  when more interaction behaviors are available.

\begin{figure}[!htb]
    \centering
    \includegraphics[width=\textwidth]{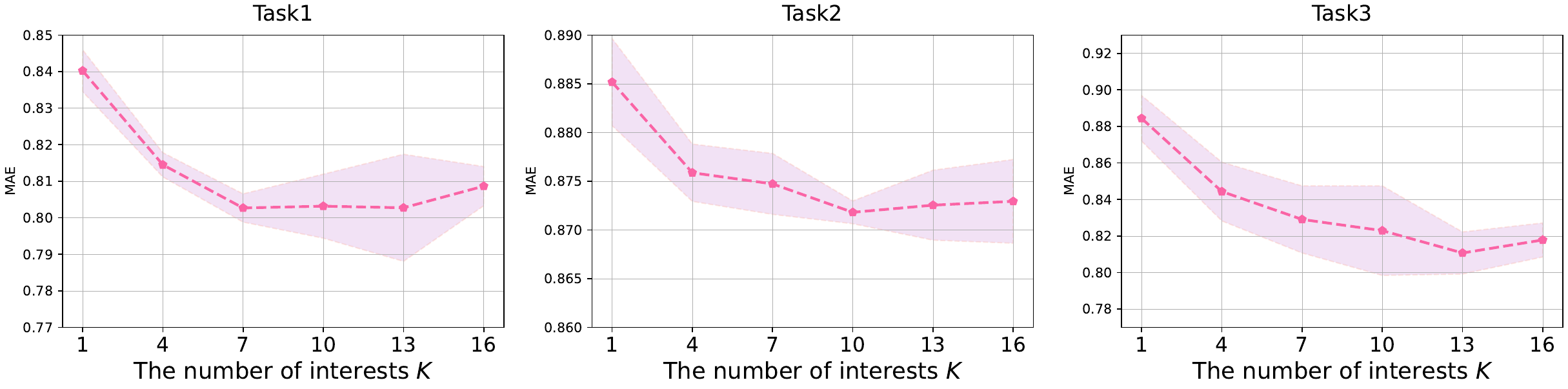}
    \caption{ Impact of the interest numbers $K$ with $\beta =20\%$, in terms of MAE.}
    \label{fig:different_K}
\end{figure}

\subsubsection{Impact of Interest Numbers $K$} 
We investigate how the number of interests, denoted as $K$, affects the performance of MIMNet.
Specifically, we vary the interest number from 1 to 16 with a step size of 3. The results are presented in Figure \ref{fig:different_K}. 
On the Task1, we can observe that the performance of MIMNet rises up first with the increment of $K$, and reaches a peak when $K$ =7.  The performance keeps stable until $K$ =13 and then it starts to decline  if $K$ becomes larger. Similar trends can also be observed on  Task2 and Task3. 
This results demonstrate that our proposed model prefer a relative large number of interest, which means employing multiple interest bridges to transfer user preference will significantly boost model performance compared to adopting a single interest bridge.
Moreover, a larger $K$ facilitates the extraction of effective user preference from different perspectives, leading to a more precise user modeling. However, a too high value of $K$ may introduce unnecessary interests, potentially degrading the performance of our model.

\subsubsection{Visualization of Multiple Interest Weight Distributions} 
\begin{figure*}[h]
    \centering
    \includegraphics[width=0.7\linewidth]{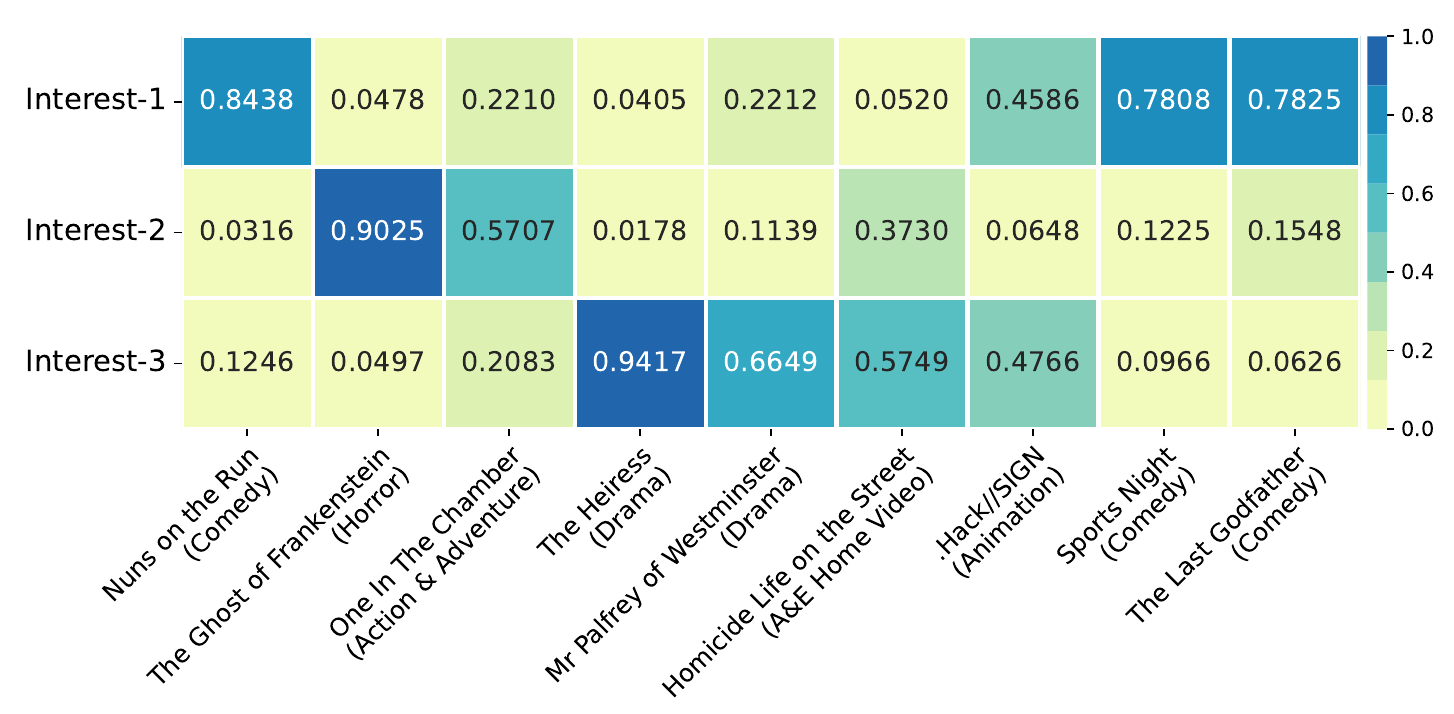} 
     \caption{ Visualization of multiple interest weight distributions of a user  in the  Task2  when $\beta=20\%$. ({Best viewed in color.})}
    \label{fig:visual-interest-user}    
\end{figure*}
To verify the effectiveness of the learned multiple user interests, we visualize the weight distribution of different interests extracted via the capsule network. 
To simplify the analysis, we only extract three interests for a sampled user, whose interacted items together their corresponding categories are ``\textit{Nuns on the Run} (Comedy), \textit{The Ghost of Frankenstein} (Horror), \textit{One In The Chamber} (Action \& Adventure), \textit{The Heiress} (Drama), \textit{Mr Palfrey of Westminster} (Drama), \textit{Homicide Life on the Street} (A\&E Home Video), \textit{.Hack//SIGN} (Animation), \textit{Sports Night}  (Comedy) and \textit{The Last Godfather}  (Comedy)''. 
Figure \ref{fig:visual-interest-user} illustrates the learned weight distribution of each item along these interests, where each row corresponds  an interest and each column corresponds an interacted item. 
We can observe that the items with same or close categories will have similar  interest weight distribution. For example, the first, eighth and ninth items which belong to the same category (i.e., ``Comedy''), all assign most of their weights on the interest-1. Similar phenomenon can also be observed on the second and third items which have  close categories (e.g., ``Horror" and ``Action \& Adventure''). 
The results indicate that our proposed model can effectively learn expressive   and decoupled multiple interests for a user, which subsequently benefits for guiding the transfer process.

\subsubsection{{Convergence Study}}
\begin{figure*}[!h]
    \centering
    \includegraphics[width=\textwidth]{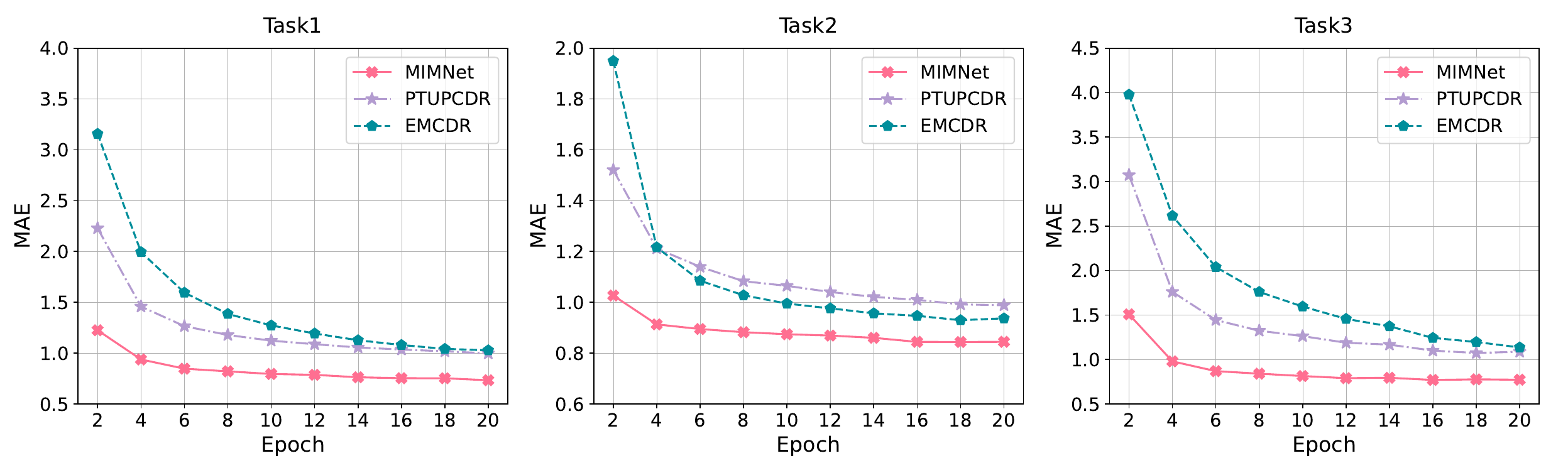}
    \caption{Convergence analysis of our proposed model  MIMNet and the other two baselines with $\beta= 20\%$, in terms of MAE.}
    \label{fig:learning-curve}
\end{figure*}
We conduct experiments to compare the convergence speed of our proposed  model MIMNet with two baselines, i.e.,  PTUPCDR and EMCDR. Figure \ref{fig:learning-curve} shows the convergence speed of different models on all three tasks with $\beta= 20\%$. 
We observe that the personalized bridge-based model PTUPCDR converges faster than the common bridge-based model EMCDR, and it also consistently  yields better performance than EMCDR at each epoch. The results demonstrate the superiority of learning a specific bridge function for each user to transfer her preference.
In addition, among all three comparing models, our proposed model demonstrates the best convergence speed. It  requires  only 10 epochs to achieve the best performance, while the two baselines generally take around 20 epochs to converge. Moreover,  the performance of MIMNet is consistently better than that of the two baselines at each epoch. The main reason is that MIMNet facilitates  the transfer of user preference in a more fine-grained manner, which leverages multi-interest bridges to transfer user embeddings from the source domain to the target domain.



\subsubsection{\textbf{Case Study.}}
\begin{figure}[!t]
    \centering
    \includegraphics[width=0.92\linewidth]{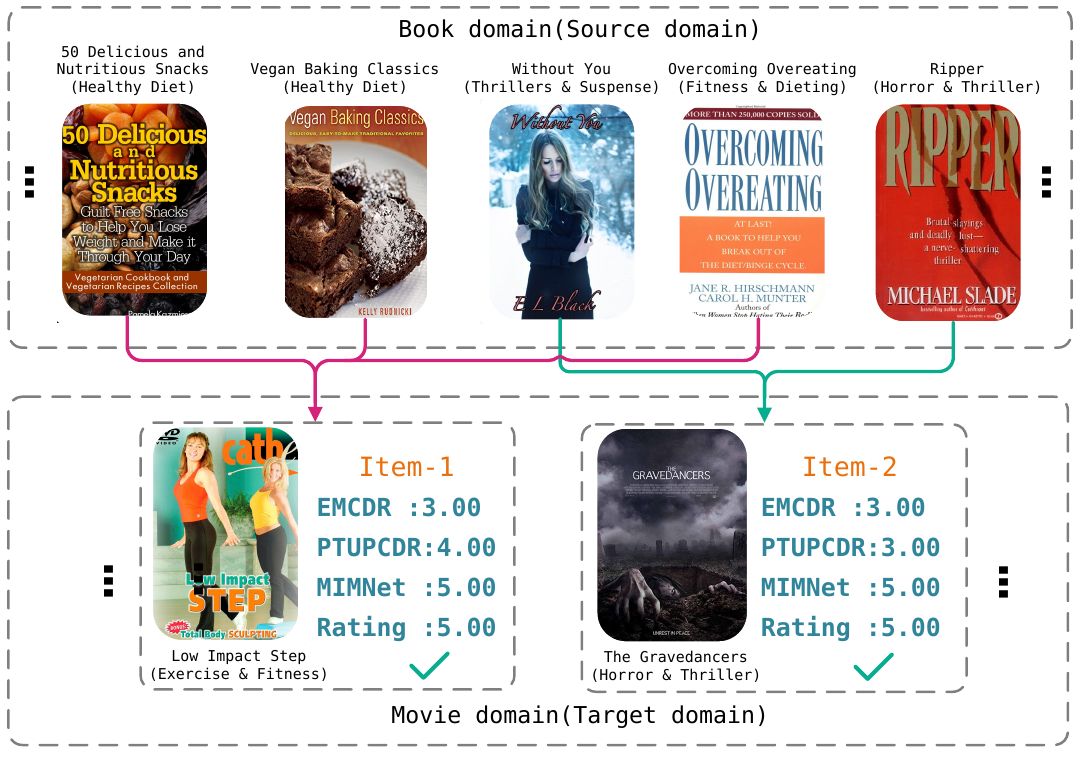}
    \caption{Case study for the Task2. 
    ({Best viewed in color.})
    }
    \label{fig:case-study}
\end{figure}

To further illustrate why our proposed method outperforms the  state-of-the-art baselines, we present a case study of a real user from Task2 in Figure \ref{fig:case-study}. 
We can observe that for the two items (i.e., \textit{Low Impact Step}  and  \textit{The Gravedancers})  interacted by the user in the target domain (i.e., Movie), both EMCDR and PTUPCDR make incorrect predictions. In contrast, our proposed method MIMNet predicts the same ratings as the real ratings of the two items, respectively. The result is attributed to that the two baselines EMCDR and PTUPCDR both transfer user preference from the source domain to the target domain in a coarse-grained manner, while ignoring the fine-grained interest-level user preference. 
To be specific, in the source domain (i.e., Book), the user interacts with five items, including  \textit{50 Delicious and Nutritious Snacks (Healthy Diet),  Vegan Baking Classics (Healthy Diet), Without You (Thrillers \& Suspense), Overcoming Overeating (Fitness \& Dieting), Ripper (Horror \& Thriller) }. Both EMCDR and PTUPCDR consider all these items together with a common bridge or a personalized bridge to transfer the user preference. However, we can see that the first, second and fourth items in Figure \ref{fig:case-study} indicate an interest in Healthy, while the remaining two items relate to an interest in Thriller. Therefore, it is more reasonable to transfer user preference  separately as conducted in MIMNet, which utilizes a multi-interest bridge  to transfer user diverse interests effectively.


\section{CONCLUSION}
This paper introduces MIMNet, a novel framework for solving the cold-start problem in cross-domain recommendation. 
To be specific, we propose a novel multi-interest meta network by utilizing dynamic routing mechanism to obtain users’ multiple interests, and generating
multi-interest bridges to transfer user embeddings from the source do-
main to the target domain. Moreover, we develop a multi-granularity target-guided attention network via exploring both fine-grained and coarse-grained target signal as a guidance of learning better transformed user representation in the target domain. 
Extensive experiments on three CDR tasks demonstrate that our proposed method can considerably outperform state-of-the-art baselines in cross-domain recommendation. Further studies verify the effectiveness of the compatibility and robustness of our model as well as the importance of each model component.

\textbf{Acknowledgments:} 
This work was supported by the National Natural Science Foundation of China [grant number 62141201];  the Natural Science Foundation of Chongqing, China [grant number CSTB2022NSCQ-MSX1672]; the Major Project of Science and Technology Research Program of Chongqing Education Commission of China [grant number KJZD-M202201102].

\bibliographystyle{elsarticle-num} 
\bibliography{main}

\begin{thebibliography}{10}
\expandafter\ifx\csname url\endcsname\relax
  \def\url#1{\texttt{#1}}\fi
\expandafter\ifx\csname urlprefix\endcsname\relax\def\urlprefix{URL }\fi
\expandafter\ifx\csname href\endcsname\relax
  \def\href#1#2{#2} \def\path#1{#1}\fi

\bibitem{zuo2023dynamic}
X.~Zuo, S.~Liang, X.~Yuan, S.~Yu, B.~Yang, Dynamic item feature modeling for rating prediction in recommender systems, Neurocomputing 549 (2023) 126412.

\bibitem{zhu2024collaborative}
Y.~Zhu, L.~Wu, Q.~Guo, L.~Hong, J.~Li, Collaborative large language model for recommender systems, in: Proceedings of the ACM on Web Conference 2024, 2024, pp. 3162--3172.

\bibitem{guo2024lgmrec}
Z.~Guo, J.~Li, G.~Li, C.~Wang, S.~Shi, B.~Ruan, Lgmrec: Local and global graph learning for multimodal recommendation, in: Proceedings of the AAAI Conference on Artificial Intelligence, Vol.~38, 2024, pp. 8454--8462.

\bibitem{wu2024personalized}
Y.~Wu, R.~Xie, Y.~Zhu, F.~Zhuang, X.~Zhang, L.~Lin, Q.~He, Personalized prompt for sequential recommendation, IEEE Transactions on Knowledge and Data Engineering (2024).

\bibitem{yu2022mdp2}
S.~Yu, Z.~Liu, S.~Wan, J.~Zheng, Z.~Li, F.~Zhou, {MDP2} forest: {A} constrained continuous multi-dimensional policy optimization approach for short-video recommendation, in: Proceedings of the 28th {ACM} {SIGKDD} Conference on Knowledge Discovery and Data Mining, 2022, pp. 2388--2398.

\bibitem{wang2023diffusion}
W.~Wang, Y.~Xu, F.~Feng, X.~Lin, X.~He, T.~Chua, Diffusion recommender model, in: Proceedings of the 46th International {ACM} {SIGIR} Conference on Research and Development in Information Retrieval, 2023, pp. 832--841.

\bibitem{bai2024labelcraft}
Y.~Bai, Y.~Zhang, J.~Lu, J.~Chang, X.~Zang, Y.~Niu, Y.~Song, F.~Feng, Labelcraft: Empowering short video recommendations with automated label crafting, in: Proceedings of the 17th ACM International Conference on Web Search and Data Mining, 2024, pp. 28--37.

\bibitem{xu2023multi}
Y.~Xu, H.~Chen, Z.~Wang, J.~Yin, Q.~Shen, D.~Wang, F.~Huang, L.~Lai, T.~Zhuang, J.~Ge, X.~Hu, Multi-factor sequential re-ranking with perception-aware diversification, in: Proceedings of the 29th {ACM} {SIGKDD} Conference on Knowledge Discovery and Data Mining, 2023, pp. 5327--5337.

\bibitem{chen2024macro}
H.~Chen, Y.~Bei, Q.~Shen, Y.~Xu, S.~Zhou, W.~Huang, F.~Huang, S.~Wang, X.~Huang, Macro graph neural networks for online billion-scale recommender systems, in: Proceedings of the ACM on Web Conference 2024, 2024, pp. 3598--3608.

\bibitem{hu2023boss}
X.~Hu, Y.~Cheng, Z.~Zheng, Y.~Wang, X.~Chi, H.~Zhu, {BOSS:} {A} bilateral occupational-suitability-aware recommender system for online recruitment, in: Proceedings of the 29th {ACM} {SIGKDD} Conference on Knowledge Discovery and Data Mining, 2023, pp. 4146--4155.

\bibitem{zheng2024mirror}
Z.~Zheng, X.~Hu, S.~Gao, H.~Zhu, H.~Xiong, Mirror: A multi-view reciprocal recommender system for online recruitment, in: Proceedings of the 47th International ACM SIGIR Conference on Research and Development in Information Retrieval, 2024, pp. 543--552.

\bibitem{cite-EMCDR}
T.~Man, H.~Shen, X.~Jin, X.~Cheng, Cross-domain recommendation: An embedding and mapping approach, in: Proceedings of the Twenty-Sixth International Joint Conference on Artificial Intelligence, 2017, pp. 2464--2470.

\bibitem{cite-DCDCSR}
F.~Zhu, Y.~Wang, C.~Chen, G.~Liu, M.~A. Orgun, J.~Wu, A deep framework for cross-domain and cross-system recommendations, in: Proceedings of the Twenty-Seventh International Joint Conference on Artificial Intelligence, 2018, pp. 3711--3717.

\bibitem{cite-SSCDR}
S.~Kang, J.~Hwang, D.~Lee, H.~Yu, Semi-supervised learning for cross-domain recommendation to cold-start users, in: Proceedings of the 28th {ACM} International Conference on Information and Knowledge Management, 2019, pp. 1563--1572.

\bibitem{cite-TMCDR}
Y.~Zhu, K.~Ge, F.~Zhuang, R.~Xie, D.~Xi, X.~Zhang, L.~Lin, Q.~He, Transfer-meta framework for cross-domain recommendation to cold-start users, in: Proceedings of the 44th International {ACM} {SIGIR} Conference on Research and Development in Information Retrieval, 2021, pp. 1813--1817.

\bibitem{cite-PTUPCDR}
Y.~Zhu, Z.~Tang, Y.~Liu, F.~Zhuang, R.~Xie, X.~Zhang, L.~Lin, Q.~He, Personalized transfer of user preferences for cross-domain recommendation, in: Proceedings of the Fifteenth ACM International Conference on Web Search and Data Mining, 2022, pp. 1507--1515.

\bibitem{sun2023reinforced}
C.~Sun, J.~Gu, B.~Hu, X.~Dong, H.~Li, L.~Cheng, L.~Mo, {REMIT:} reinforced multi-interest transfer for cross-domain recommendation, in: Proceedings of the 37th {AAAI} Conference on Artificial Intelligence, 2023, pp. 9900--9908.

\bibitem{cite-CMF}
A.~P. Singh, G.~J. Gordon, Relational learning via collective matrix factorization, in: Proceedings of the 14th {ACM} {SIGKDD} International Conference on Knowledge Discovery and Data Mining, 2008, pp. 650--658.

\bibitem{cite-CST}
W.~Pan, E.~W. Xiang, N.~N. Liu, Q.~Yang, Transfer learning in collaborative filtering for sparsity reduction, in: M.~Fox, D.~Poole (Eds.), Proceedings of the Twenty-Fourth {AAAI} Conference on Artificial Intelligence, {AAAI} Press, 2010, pp. 230--235.

\bibitem{cite-CoNet}
G.~Hu, Y.~Zhang, Q.~Yang, Conet: Collaborative cross networks for cross-domain recommendation, in: Proceedings of the 27th {ACM} International Conference on Information and Knowledge Management, 2018, pp. 667--676.

\bibitem{cite-MINDTL}
M.~He, J.~Zhang, P.~Yang, K.~Yao, Robust transfer learning for cross-domain collaborative filtering using multiple rating patterns approximation, in: Proceedings of the Eleventh {ACM} International Conference on Web Search and Data Mining, 2018, pp. 225--233.

\bibitem{cite-DDTCDR}
P.~Li, A.~Tuzhilin, {DDTCDR:} deep dual transfer cross domain recommendation, in: Proceedings of the Thirteenth {ACM} International Conference on Web Search and Data Mining, 2020, pp. 331--339.

\bibitem{jenatton2012latent}
R.~Jenatton, N.~L. Roux, A.~Bordes, G.~Obozinski, A latent factor model for highly multi-relational data, in: Proceedings of the 26th Annual Conference on Neural Information Processing Systems, 2012, pp. 3176--3184.

\bibitem{sabour2017dynamic}
S.~Sabour, N.~Frosst, G.~E. Hinton, Dynamic routing between capsules, in: Proceedings of the 31st Advances in Neural Information Processing, 2017, pp. 3856--3866.

\bibitem{li2019multi}
C.~Li, Z.~Liu, M.~Wu, Y.~Xu, H.~Zhao, P.~Huang, G.~Kang, Q.~Chen, W.~Li, D.~L. Lee, Multi-interest network with dynamic routing for recommendation at tmall, in: Proceedings of the 28th {ACM} International Conference on Information and Knowledge Management, 2019, pp. 2615--2623.

\bibitem{zhu2021transfer}
Y.~Zhu, K.~Ge, F.~Zhuang, R.~Xie, D.~Xi, X.~Zhang, L.~Lin, Q.~He, Transfer-meta framework for cross-domain recommendation to cold-start users, in: Proceedings of the 44th International {ACM} {SIGIR} Conference on Research and Development in Information Retrieval, 2021, pp. 1813--1817.

\bibitem{he2017neural}
X.~He, L.~Liao, H.~Zhang, L.~Nie, X.~Hu, T.~Chua, Neural collaborative filtering, in: Proceedings of the 26th International Conference on World Wide Web, 2017, pp. 173--182.

\bibitem{covington2016deep}
P.~Covington, J.~Adams, E.~Sargin, Deep neural networks for youtube recommendations, in: Proceedings of the 10th {ACM} Conference on Recommender Systems, 2016, pp. 191--198.

\end{thebibliography}

\end{document}